# GRAPH THEORY IN THE CLASSIFICATION OF INFORMATION SYSTEMS

Daniel ZENTAI
Óbuda University
Doctoral School of Safety and Security Science
zentai.daniel@bgk.uni-obuda.hu

***Abstract:*** *Risk classification plays an important role in many regulations and standards. However, a general method that provides an optimal classification has not been proposed yet. Also, the criteria of optimality are not defined in these regulations. In this work, we will propose a mathematical model that is sufficient to describe this problem, and we also propose an algorithm that classifies graph vertices based on their risk value in polynomial time.*

## 1. INTRODUCTION

From a mathematical point of view, the process of defending and attacking an information system, is a (not necessarily zero-sum) game between an attacker and a defender. The payoff of the attacker is the profit he can make with the attack, and the payoff of the defender is the value of the information he protects. As the defender we have to be aware of the risks and values of the components of our system, in other words, we have to assign our components to classes based on their risk values.

The necessity of classifying information systems appears in some regulations and standards for a long time. In Hungary, this also appears in the Act on the Electronic Information Security of Central and Local Government Agencies (Act L of 2013/Information Security Act). In this work, we will focus on an efficient algorithm that will give us the best possible (however, still not optimal) classification.

Usually, information systems form a network, where the nodes are the components, and two nodes are adjacent if there exists a direct communication channel between them.

The most efficient tool for modeling communication networks is graph theory. A graph is a finite set of vertices, connected with edges. Graph theory is a widely used and efficient way to analyze various properties of networks, such as a railway network, a communication network, a computer network, or even the network of neurons in the human brain.

The most natural way to measure the robustness, of a graph, is multiple connectivity [1]. Suppose that a graph $G = (V, E)$ has some prescribed local connectivity value amongst the vertices. The aim of this paper is to propose an efficient algorithm to classify the vertices of the graph into separate clusters, as optimally as it is possible. A polynomial-time algorithm is proposed in this work, which classifies the vertices of the graph with an approximation factor of 2.

## 2. PRELIMINARIES

In this section, some basic definitions and theorems of graph theory will be described, which are needed to understand the following results.

Let $V = \{v_1, v_2, \ldots, v_n\}$ be a finite set, and E is a set of 2-element subsets of V, i.e. $E \subseteq \binom{V}{2}$. The structure G = (V,E) formed by these two sets is called a finite graph. The elements of the set V = V(G) are the vertices, and the elements of the set E = E(G) are the edges of the graph. The vertices v, w ∈ V are neighboring, if they are connected with an edge, i.e. {v,w} ∈ E. The set of the neighbors of some vertex v ∈ V is denoted by $N(v)$. The degree of a vertex a v ∈ V is the number of the neighbors of v, and it is denoted by d(v), i.e. $d(v)=|N(v)|$.

A walk in a graph is an alternating sequence of vertices and edges, $v_0 e_1 v_1 \ldots e_k v_k$, where every edge is connected to the vertex right before, and after the edge. i.e. $e_i = \{v_{i-1}, v_i\}$. A trail is a walk, where every edge is covered at most once. A path is a trail, where every vertex is covered at most once.

The graph G = (V,E) is said to be connected, if every vertex is reachable with edges from every other vertex. More precisely, G is connected, if there exists a path from u to v for every u,v ∈ V vertices. Let G = (V,E) a (not necessarily connected) graph. A subset of the vertices C ⊆ V(G) is a connected component if there is a path from u to v for every u,v ∈ C, but there are no paths that lead outside of C, namely, there is no path from u to v, if u ∈ C, and u,v ∈ V \ C. The number of connected components of G is denoted by c(G), therefore a graph is connected if and only if c(G) = 1.

The most natural way to measure the robustness of a graph is to calculate the connectivity number of the graph.

*Definition*: A graph G = (V,E) is *k*-connected, if $|V(G)| \geq k+1$, G is connected, and it remains connected if we remove an arbitrary set of fewer than *k* vertices with the adjacent edges. The greatest value *k* for which G is *k*-connected, is the connectivity number of G, and is denoted by κ(G).

*Definition*: A graph G = (V,E) is *k*-edge-connected, if G is connected, and it remains connected if we remove an arbitrary set of fewer than *k* edges. The greatest value *k* for which G is *k*-edge-connected, is the connectivity number of G, and is denoted by λ(G).

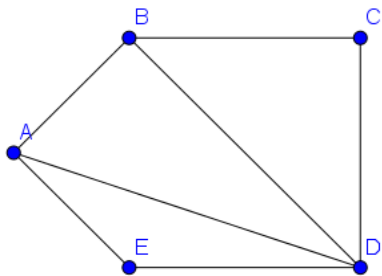

Fig. 2. A graph which is 2-connected and also 2-edge-connected

The edge connectivity models the robustness of the graph against an attack on the connections between the communicating units. On the other hand, vertex connectivity models the robustness against an attack on the communicating units themselves. There are well-known efficient algorithms to calculate the connectivity, or edge-connectivity number of a graph [2], but it is not efficient to perform such a calculation by the definition of multiple connectivity. Therefore, it is reasonable to recall some equivalent reformulation of multiple connectivity. These theorems are immediate consequences of Menger's theorem.

*Theorem*: The following statements are equivalent:
- The G = (V,E) graph is *k*-edge-connected.
- For all pair of vertices u,v ∈ V(G), there exists *k* number of edge-disjoint u-v paths, i.e. *k* paths such that none of them have a common edge.

*Theorem*: The following statements are equivalent:
- The G = (V,E) graph is *k*-edge-connected.
- For all pair of vertices u,v ∈ V(G), there exists *k* number of disjoint u-v paths, i.e. *k* paths such that none of them have a common vertex, except the start and the end vertex.

The local connectivity of graphs is defined as follows.

*Definition*: Let G = (V,E) a graph, and u,v ∈ V(G) arbitrary vertices. The local edge-connectivity between the vertices u and v is the maximum number of edge-disjoint paths between u and v, and it is denoted by λ(u,v).

*Definition*: Let G = (V,E) a graph, and u,v ∈ V(G) arbitrary vertices. The local connectivity between the vertices u and v is the maximum number of disjoint paths between u and v, and it is denoted by κ(u,v).

It is a well-known fact, that the edge-connectivity of a graph is nothing else, but the minimal value amongst the local edge-connectivity numbers. Similarly, the connectivity of a graph is the minimal value amongst the local connectivity number. More formally, λ(G) = min{λ(u,v) | u,v ∈ V(G)}, and κ(G) = min{κ(u,v) | u,v ∈ V(G)}.

## 3. RISK CLASSIFICATION

It is possible now to ask ourselves the following question. If there are different local connectivity requirements between the vertices of a graph, how can the vertices be classified in a way, that every vertices in the same class have a similar connectivity requirement? In other words, how can we assign our information systems to risk classes in an optimal way. In this case, optimality means that the maximum distance between elements of different sets should be minimal.

Classification is a well-known problem in data mining [5]. But before the classification, it is necessary to define some similarity measure on the vertices, and then put the vertices into different classes based on their similarity. In our case, the similarity measure will be a risk-based function.

Formally classification is the following problem. Split the vertices of a graph G = (V,E) into partitions Γ=($C_1$,…,$C_k$) for some fixed integer *k*, such that $C_1$∪…∪$C_k$ = V and the $C_i$ sets are pairwise disjoints. Additionally, the goal is to minimize the maximum distance between the vertices inside each cluster for some distance function D. More formally we have to minimize the value M(Γ)=max{D(u,v) | u, v ∈ C, C is a cluster},

Unfortunately, this problem was proven to be NP-complete [4], which means we cannot find the optimal solution in polynomial time. However, it is possible to create a 2-approximation algorithm, i.e. an algorithm, that finds a set of clusters, in which the maximum distance between the vertices inside each cluster is at most 2 times the value of the optimal clustering.

Now the definition of the distance function is as follows. To do so, the common mathematical definition of risk is used in this section, with some modification, that is more applicable with a graph theoretic approach.

Let $r_0 : V(G) \to \mathbb{R}_+$ be the function

$$r_0(v) = \sum_{t \in T(v)} p(t)d(t)$$

Here $r_0(v)$ is an initial risk value corresponding to the vertex v, T(v) is the set of possible threats against v, p(t) is the probability of the threat t, and d(t) is the potential value of the damage that threat t can cause. This is the most natural definition of risk, and this definition is used widely in the literature [3]. Unfortunately, this initial risk value is not the most realistic model of our problem, since the connections between the vertices are note encoded in this formula. However, it is possible, that if two vertices have the same initial risk value, but one of them is connected to a set of vertices with a high risk value, then it is logical to evaluate this vertex as a riskier vertex than the other one. This situation is demonstrated in the following picture. Suppose that the vertices u and v have the same initial risk i.e. they are vulnerable against the same attacks, and their damage occurs the same amount of financial loss. However, N(v) contains vertices with risk value 1, but N(u) also contains a vertex with risk value 10. It is a realistic assumption, that in u should have a higher risk value, since removing u from the graph may affect the local connectivity of a vertex with higher risk value.

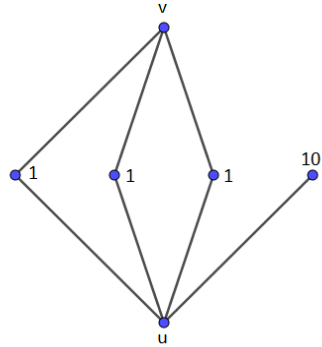

Fig. 3.

Let $r : V(G) \to \mathbb{R}+$ denote the modified risk value. It is defined as follows.

$$r(v) = r_0(v) + f(v)$$

where f(v) denotes the average initial risk value of the vertices in N(v). More formally

$$f(v) = \frac{1}{d(v)} \sum_{u \in N(v)} r_0(u)$$

Now it is possible to define the distance of two arbitrary vertices D(u,v), as the difference between the risk value of the vertices u and v, i.e. D(u,v) = |r(u) – r(v)|. Naturally, if this distance is small, then u and v should be in the same cluster.

To apply any clustering algorithm, first it is necessary to show that D is indeed a distance function. Mathematically, a distance function $g : A \times A \to \mathbb{R}$ on a set A is a function with the following properties:

- non-negativity:
  $g(x,y) \geq 0$ for all x, y

- $g(x,y) = 0$, if $x=y$
- symmetry:
  $g(x,y) = g(y,x)$ for all $x,y$
- triangle-inequality:
  $g(x,y) \leq g(x,z) + g(z,y)$ for all $x,y,z$

Now check these properties in the case of the aforementioned function $D : V(G) \times V(G) \to \mathbb{R}$.

*Claim*: If u and v are arbitrary vertices, then $D(u,v) = |r(u) - r(v)|$ is a distance function.

Proof:

- non-negativity:
  $D(u,v) \geq |r(u) - r(v)| \geq 0$ for all u, v, since $D(u,v)$ is the absolute value of a real number
- If u=v, then $D(u,v) = D(u,u) = |r(u) - r(u)| = 0$
- symmetry:
  $D(u,v) = |r(u) - r(v)| = |-(r(v) - r(u))| = |-1||r(v) - r(u)| = |r(v) - r(u)| = D(v,u)$
- triangle-inequality:
  $D(u,v) = |r(u) - r(v)| = |r(u) - r(w) + r(w) - r(v)| \leq |r(u) - r(w)| + |r(w) - r(v)| = D(u,w) + D(w,v)$ for all u, v, w vertices

Besides of the distance of vertices, the distance of clusters can be defined as follows. If C, C' $\subseteq$ V are arbitrary clusters, then, the distance of C and C' is $D(C,C') = \min\{D(u,v) \mid u \in C, v \in C'\}$.

Our proposed algorithm works as follows.
1) Initially let $C_1=\{v_1, v_2, \ldots, v_n\}$ the only cluster that contains all vertices.
2) At a general step of the algorithm suppose that we have created *i* clusters so far. Pick an arbitrary vertex from each clusters $\{C_1, C_2, \ldots, C_i\}$, and label this vertex as the center of the cluster.
3) Find a vertex *w*, whose distance from its cluster's center is maximal. Let *w* be the center of the new cluster $C_{i+1}$
4) Stop, if only k clusters have remained.

As we mentioned in the previous section, this algorithm does not always give the optimal clustering as its output. However, it is provable, that the algorithm always provides a clustering with maximal distance at most 2 times the optimal distance in polynomial time.

## 5. CONCLUSION

In this work, we proposed an algorithm for clustering graph vertices based on their risk value. Since the optimal classification is NP-complete, we proposed an algorithm that gives an output that is at most 2-times worse than the optimal solution. The algorithm runs in polynomial time, and it has the best possible approximation factor amongst algorithms with this efficiency.

## 4. ACKNOWLEDGEMENT

Supported by the ÚNKP-19-3 New National Excellence Program of the Ministry for Innovation and Technology.

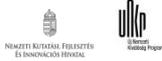